\documentclass{llncs}

\usepackage{amssymb}
\usepackage{color}
\usepackage{graphicx}
\usepackage{url}

\graphicspath{{figs/}}

\newcommand{\codename}[1]{\texttt{#1}}

\begin{document}

\title{An Extensible Timing Infrastructure for Adaptive Large-scale
  Applications}

\author{Dylan~Stark\inst{1} \and Gabrielle~Allen\inst{1} \and
  Tom~Goodale\inst{1} \and Thomas~Radke\inst{2} \and
  Erik~Schnetter\inst{1}}
\institute{Center for Computation \& Technology, Louisiana State
  University,\\
  216 Johnston Hall, Baton Rouge, LA 70803, USA\\
  \url{http://www.cct.lsu.edu/},
  \email{\textless dstark@cct.lsu.edu\textgreater}
  \and Max Planck Institute for Gravitational Physics (Albert Einstein
  Institute),\\
  Am M\"uhlenberg 1, D-14476 Golm, Germany}

\maketitle

\begin{abstract}
  Real-time access to accurate and reliable timing information is
  necessary to profile scientific applications, and crucial as
  simulations become increasingly complex, adaptive, and large-scale.
  The Cactus Framework provides flexible and extensible capabilities
  for timing information through a well designed infrastructure and
  timing API\@.  Applications built with Cactus automatically gain
  access to built-in timers, such as \codename{gettimeofday} and
  \codename{getrusage}, system-specific hardware clocks, and
  high-level interfaces such as PAPI\@.  We describe the Cactus timer
  interface, its motivation, and its implementation.  We then
  demonstrate how this timing information can be used by an example
  scientific application to profile itself, and to dynamically adapt
  itself to a changing environment at run time.
\end{abstract}

\section{Introduction}

Profiling has long been an important part of application development. 
 In the early days profiling was
restricted to overall performance metrics such as wall-clock time for
a particular calculation or routine, and optimisation was often limited to
finding better algorithms --- ones which would take fewer operations.
Today it is possible to access hardware counters which
give developers information on memory metrics such as cache behaviour
and floating point performance, and many tools such as SGI Speedshop,
Intel's VTUNE, or the Sun Studio Performance Analyzer are available
which can provide a complete performance profile of an application
down to individual source lines.  These tools are excellent for
application developers tuning their codes, but are not useful for
adaptive tuning by applications themselves, and further are limited to 
particular platforms or operating systems.  Self-tuning of
applications is becoming increasingly important in today's world of
massively networked, dynamic data-driven applications~\cite{CS_dddasworkshop_06}
and the Grid computing and 
peta-scale
applications of tomorrow. 

For modern applications it is necessary to have a programming API which
allows the application to query and analyze its own performance
characteristics on-the-fly.  It must be easy to create caliper points
between which to measure performance and to query them, and it must be
possible to access the wide range of different metrics available on
modern hardware.  In this paper we describe the approach taken within
the Cactus framework.  Cactus provides a rich timing API which can be
used with basic timing metrics such as wall clock or user CPU time, or
more sophisticated metrics available by plugging in libraries such as
the Performance API (PAPI)~\cite{CS_papi_web} developed at the University of Tennessee.

The \emph{Cactus
  Framework}~\cite{CS_cactus_web,CS_Goodale02a,CS_cactususersguide_nourl}
is
an open source, modular, highly portable programming environment for
collaborative high performance computing.  Cactus has a generic
parallel toolkit for scientific computing with modules providing
parallel drivers, coordinates, boundary conditions, elliptic solvers,
interpolators, reduction operators, and efficient I/O in different
data formats. Also, 
generic interface definitions (e.g.\ an abstract elliptic
solver API) make it possible to use external packages 
and improved modules, which
are immediately available to users of the abstract interface.

Although Cactus originated in the numerical relativity community, it is now used as 
an enabling HPC framework for applications in many disciplines including computational 
fluid dynamics, coastal modeling, astrophysics, and quantum gravity. Cactus has also 
been a driving application for computer science research, particularly in Grid and distributed 
computing. 
For example, the so-called ``Cactus-Worm'' application~\cite{CS_Allen01a_nourl}  used contracts based on runtime performance to trigger migration of an astrophysics application across distributed Grid resources. Another Cactus application used MPICH-G2 to distribute a single simulation across multiple machines connected by wide area networks. Using adaptive algorithms to tune communication 
patterns to available bandwidth, this application showed good overall scaling~\cite{CS_Allen01d_nourl}. 

By using abstract interfaces 
for accessing meta-information about the system state on which
it is running, Cactus enables an application to be aware of its
surroundings in a very portable, system-independent manner.  This
allows users to easily implement and experiment with dynamic
scenarios, such as responding to increased delays in disk I/O times,
adapting algorithmic parameters to changes in an AMR (adaptive mesh
refinement) grid hierarchy, or postponing analysis methods from
in-line to a post-processing step.

In this paper, we describe the design and implementation of the Cactus
timing infrastructure. Sec.~\ref{cti} covers the timing infrastructure and clock API\@.  Sec.~\ref{uc}
discusses how it can be used in different applications scenarios, including a
 new use to adaptively control
checkpointing intervals for large scale simulations.
In Sec.~\ref{ea} the results of a case study for the checkpointing scenario are presented, and
 Sec.~\ref{rwc}  compares
the Cactus timing infrastructure with other packages and libraries, and explains the benefits of profiling 
within the application code.

\section{Cactus Timing Infrastructure}
\label{cti}

Code using the Cactus framework is divided into modules, or components,
called \emph{thorns}. Each thorn declares an interface to Cactus
and schedules a number of routines.
Cactus controls the execution of  these routines,
providing a natural place to put caliper points to time 
routines.  The presence of timers in the scheduling mechanism obviates
the need for developers to place explicit timers in code and
allows any user or other routine to obtain timing statistics for
any routine used in a particular simulation by querying the internal timer
database.

Cactus provides a generic and extensible \emph{timing infrastructure}.
This infrastructure allows the code to access timers such as
\codename{gettimeofday}
and \codename{getrusage}, system-specific hardware clocks, and
high-level interfaces such as PAPI, all in a portable
manner. This timing information can be accessed
programmatically at runtime
through the Cactus timing API and made available to the user via
online application monitoring interfaces integrated in Cactus.
It is also logged semi-automatically for post-mortem review.

\begin{figure}[tbp]
  \includegraphics[width=0.63\textwidth]{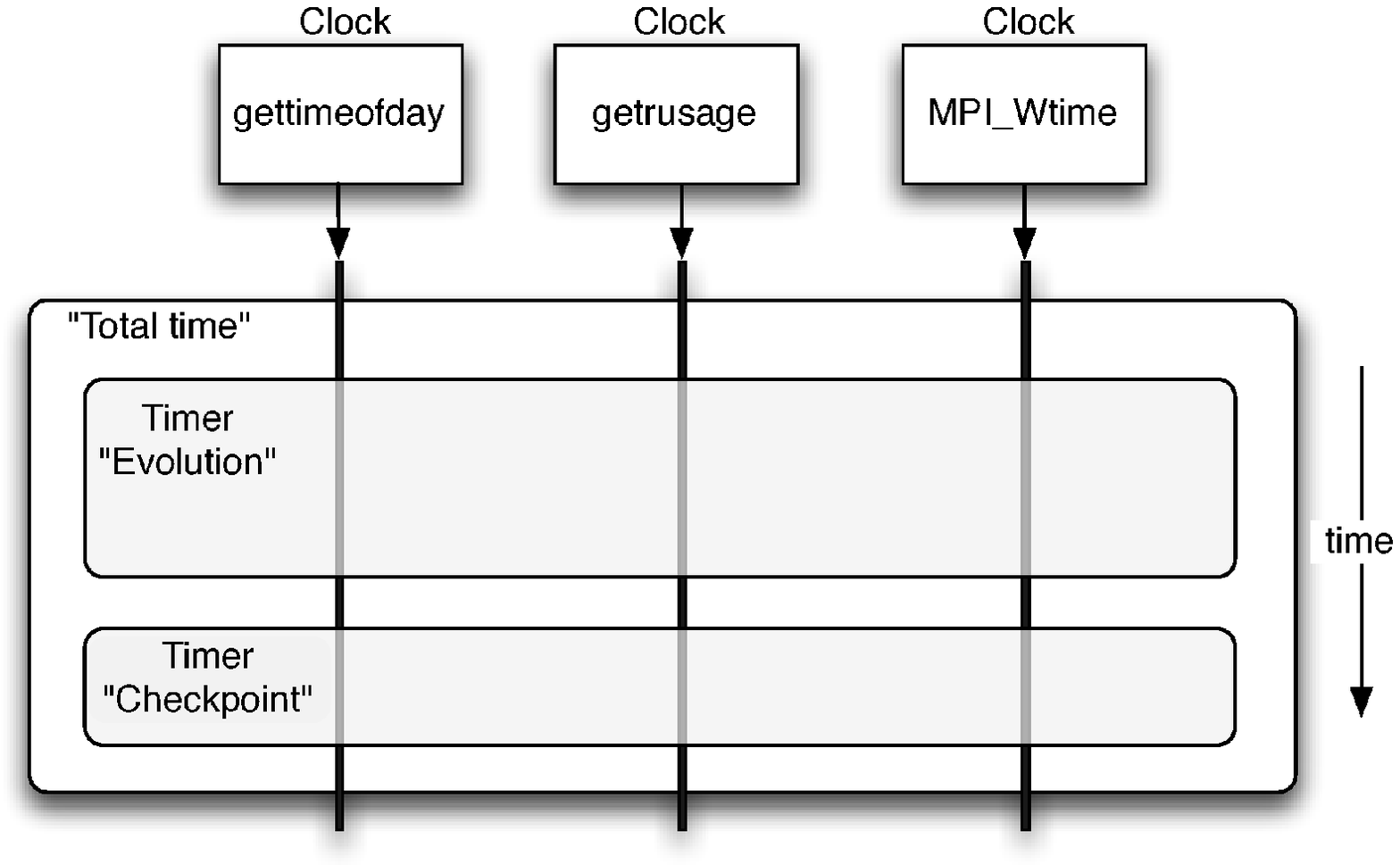}
  \hfill
  \includegraphics[width=0.35\textwidth]{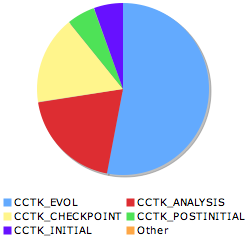}
  \caption{\emph{Left:} The relationship between timers and clocks in
    Cactus.  Clocks are low-level entities representing e.g.\ hardware
    counters, timers are used by application code.  \emph{Right:}
    Example wall time distribution onto different stages of a
    simulation.}
  \label{fig:timers-and-clocks}
\end{figure}

The Cactus timing infrastructure consists of two core concepts:
\emph{timers}, which are used to place caliper points around sections
of code and can be switched on and off or reset, and \emph{clocks},
which provide the actual timing measures, such as wall-clock time or
number of floating point operations.
Figure~\ref{fig:timers-and-clocks} shows the relationship between
clocks and timers.
Querying a timer returns the
timing results for all clocks associated with that timer.  Clocks
themselves can be registered with the timing infrastructure using
Cactus's standard registration techniques and thus can be provided by
a thorn.  This provides an extensible mechanism by which extra clocks,
and hence timing metrics, can be used with no modification to any of
the existing timing code.


Cactus thorns offer a wide variety of clocks.
Some of these clocks are only available on certain architectures, or
if certain libraries have been installed.  A new clock can be easily
added by providing callback functions to create, destroy, start, stop,
read out, and reset the clock.  Clocks are not restricted to measure
time; they can measure any kind of event, e.g.\ discrete events
such as cache misses, I/O failures, or network packet losses.
Table~\ref{tab:available-clocks} lists clocks which are currently
available.  Table~\ref{tab:clock-api} describes the Cactus clock
API\@. Clocks are usually not used directly; they are instead
encapsulated in timers.

\begin{table}[tbp]
  \begin{centering}
    \begin{tabular}{l|ll}\hline
      Clock name & Unit & Description \\\hline
      gettimeofday & sec & UNIX wall time \\
      getrusage & sec & UNIX system time \\
      MPI\_Wtime & sec & Wall time \\
      PAPI counts & & Many hardware counters, e.g.\ instructions or Flop \\
      rdtsc & & Intel CPU time stamp counter \\\hline
    \end{tabular}\\
  \end{centering}
  \vspace{2ex}
  \caption{Available clocks in Cactus.  Some of these clocks are only
    available on certain architectures, or if certain libraries have
    been installed.}
  \label{tab:available-clocks}
\end{table}

\begin{table}[tbp]
  \vspace{2ex}
  \begin{centering}
    \begin{tabular}{l|l}\hline
      Function & Description \\\hline
      create & Create a new clock, returning a pointer to it \\
      destroy & Destroy the clock \\
      start & Start this clock \\
      stop & Stop this clock \\
      reset & Reset this clock, i.e., set the accumulated time to zero
      \\
      get & Get the clock's values \\
      set & Set the clock's values \\\hline
    \end{tabular}\\
  \end{centering}
  \vspace{2ex}

  \caption{Cactus clock API\@.
  A \emph{clock} is an object which measures certain events.  Several
  clocks of the same kind can exist and can be running at the same
  time, measuring potentially overlapping durations.
  Clock can measure several values at the same time, e.g.\ multiple
  PAPI counters.
  Clocks are not meant to be called by user thorns (although this is
  of course possible); instead, clocks are encapsulated in timers.
  See also Table~\ref{tab:timer-example}.}

  \label{tab:clock-api}
\end{table}

The Cactus timing API is the interface which can be used to time or
profile events or regions of code. Timers are usually created at
startup time (or the first time a routine is entered), and they are
started and stopped before and after the events that should be
measured.  The values of the clocks associated with a timer can be output explicitly using timer calls, or using a Cactus functionality that outputs all
existing timers periodically to a log file.  
Table~\ref{tab:timer-example} gives an
example of using timers, the complete API is
described in the reference manual~\cite{CS_cactus_web}.

\begin{table}[tbp]
{\small
  \begin{verbatim}
/* Create timer */
static int handle = -1;
if (handle < 0) {
  handle = CCTK_TimerCreate ("Poisson: Evaluate residual");
  if (handle < 0) CCTK_WARN (CCTK_WARN_ABORT, "Could not create timer");
}
... other code ...
CCTK_TimerStartI (handle); /* Start timer */
... evaluate residual ...
CCTK_StopTimerI (handle); /* Stop timer */
... other code ...
CCTK_TimerPrintDataI (handle, -1); /* Output all clocks of this timer */
  \end{verbatim}
  }
  \caption{Example source code using Cactus timers, illustrating how a
    timer is created, started, stopped, and printed.}
  \label{tab:timer-example}
\end{table}

The accuracy of the timing information is obviously limited by the
accuracy of the underlying clocks.  Many clocks have accuracies
measured in microseconds, and are hence not suitable for profiling
very short events or routines.  Other clocks, such as e.g.\
\codename{rdtsc}, have nanosecond resolution and can measure with a
very fine granularity.  One has to keep in mind that measuring time
changes the instruction flow through the CPU, often acting as
barriers, so that it is impossible to measure with
sub-nanosecond accuracy on today's CPU architectures.

The Cactus timer interface is a high performance interface.  Creating
and destroying timers typically requires allocating and freeing
memory, so this should not be done in inner loops.  Starting and
stopping timers is as efficient as the underlying clocks implement it,
plus overhead from indirect function calls.  (The clocks' routines are
called via function pointers.)

\section{Use Cases}
\label{uc}

In this section we present use cases for application
self-profiling.  We give two examples describing the current use of the
timing infrastructure for automated report generation and the use of
profiling information to guide adaptive control of applications.
We also suggest some other possible application scenarios 
which are possible given the above timing infrastructure.

\subsection{Timer Report}
\label{tr}

Cactus automatically sets up timers for each scheduled routine, as
described in Sec.~\ref{cti}.  The information from these
timers is dynamically available to the application through the timer API\@.  This
information is used to provide details about application
performance while the simulation is running, 
reporting it e.g.\ to standard output, via a web-accessible HTTP
interface\footnote{See
  \url{http://cactus.cct.lsu.edu:5555/TimerInfo/index.html} for timing
  information for the perpetual Cactus demonstration run}, or
via log files.  The same mechanism can also be used to
influence the behaviour of the application, allowing it to adapt
itself to changes in the simulation or the environment.

Timer reports are generated for any Cactus application by setting
the parameter \codename{Cactus::print\_timing\_info="full"}.
Figure~\ref{fig_timer_report} shows part of such a report
for one of the runs of the use case presented below.
The information in the report is collected by querying the timers
periodically.  In this case, the two clocks available
to the simulation were \codename{gettimeofday} and
\codename{getrusage}.
Figure~\ref{fig:timers-and-clocks} shows a graphical representation of
such a report.

\begin{figure}[tbp]
  {\scriptsize
\begin{verbatim}
Thorn          | Scheduled routine in time bin    | gettimeofday [secs] | getrusage [secs] 
===========================================================================================
CarpetIOHDF5   | Evolution checkpoint routine     |         79.76328000 |      13.66692200 
-------------------------------------------------------------------------------------------
               | Total time for CCTK_CHECKPOINT   |         79.76328000 |      13.66692200 
===========================================================================================
AdaptCheck     | Adaptive checkpointing startup   |          0.00001300 |       0.00000000 
BSSN_MoL       | Register provided slicings       |          0.00000700 |       0.00000000 
===========================================================================================
               | Total time for simulation        |       1417.13730900 |    1305.43354400 
===========================================================================================
\end{verbatim}}
\vspace{-3ex}
  \caption{Part of the standard timer report available for any Cactus
    application by setting a simple parameter.  This report shows the
    time spent in some scheduled routines.}
  \label{fig_timer_report}
\end{figure}

\subsection{Adaptive Checkpointing}
\label{ac}

Checkpointing is often used by applications deployed on clusters and
supercomputers to provide protection again hardware and software
failures, to allow for simulations which require longer run times than
available on batch queues, and more recently to enable different
dynamic grid computing scenarios.  Cactus provides application-level
checkpointing which saves a snapshot of the running simulation by
writing to file all active grid variables, parameters and other state
information.  The checkpoint file uses a platform independent file
format, and the run can be restarted either on the same machine or on
a completely different architecture using a different number of
processors.

The current checkpointing mechanism in Cactus allows for checkpointing
after initial data generation, periodic checkpointing based on
iteration count, and checkpointing on termination.  We developed a new
thorn \codename{AdaptCheck} which dynamically controls the
checkpointing characteristics of a Cactus application using real-time
profiling timing information provided through the timer
infrastructure.

Writing a checkpoint file for a simulation can take a relatively long
time, depending on the number of state variables to be saved, the file
system characteristics, and the efficiency of I/O layer.  The time
needed can also vary over the lifetime of a simulation.  For example,
when using adaptive mesh refinement, the amount of data to be stored
varies with the number of refinement levels, which itself depends on
dynamic quantities such as the truncation error.

Assuming that the run is allocated some fixed amount of wall time for
which it can use a resource, a growing checkpoint time would
necessarily take away from the actual time spent on the problem.
\codename{AdaptCheck} allows the user to specify the maximum
percentage of a time the simulation should spend checkpointing in
order to best use the fixed amount of time available on some resource.
This is a weak upper bound, which means that the thorn guarantees that
no checkpoint will be performed if the current percentage of time
spent checkpointing is above the specified level, but it does not
guarantee that a checkpoint will not occur which will result in the
percentage of time being higher than the specified level.

The quality of the fault tolerance provided by checkpointing depends
on the frequency of the snapshots.  Cactus currently allows the user
to specify a checkpointing interval in terms of iterations,
independent of the runtime performance of the simulation.  Adaptive
checkpointing in the manner described above could result in long
periods of time without checkpointing.  In order to prevent this,
\codename{AdaptCheck} also respects an upper bound on the length of
wall time a simulation will progress without checkpointing.  This
guarantees that checkpoints will be generated with some regularity,
with respect to wall time.

The current implementation of \codename{AdaptCheck} uses the
\codename{gettimeofday} clock for measuring both the simulation and
checkpointing durations.  This will be extended to allow for the use
of other user-specified clocks.  We also plan to incorporate a better
prediction for the time required for the next checkpoint.  This will
then be used to remain closer to the user-specified maximum percentage
of wall time.  It will also make final checkpoints reliable, which
have to be finished before the queue time is used up.
We present in Sec.~\ref{ea} some results from tests using
\codename{AdaptCheck} to control checkpointing for an AMR code.

\subsection{Future Scenarios}

The previous examples illustrate two scenarios where application-side real-time profiling is used by large scale applications. Taking advantage of the flexible, well-designed timing infrastructure in Cactus, many other uses are planned. 

Building on the basic timing report mechanisms described in Sec.~\ref{tr}, more advanced and informative
reports can generated, for example with the web interface providing graphical interpretation of 
results, or automated documents could be produced in a readable format that can be easily interpreted. 

The technique for adaptive checkpointing can be applied to Cactus analysis thorns, whose methods are called only when output  is required. As with checkpointing, it is usual to output at regular iteration intervals, a more effective mechanism would involve choosing the output  frequency dynamically based both on user requirements and performance for a particular analysis method. 

In Grid computing, as new capabilities become available on production resources, taking advantage of application-oriented APIs such as the Simple API for Grid Applications~\cite{goodale05:saga}, previously prototyped scenarios such as simulation migration, adaptive distributed simulations, task spawning, 
will become more regularly used. Accurate information from applications will be needed to make decisions about when and how to use such services. As described in Sec.~\ref{rwc}, current profiling services rely on a remote service discovering and interpreting information from applications, however 
we believe scenarios will be more powerful and reliable when closely coupled in the application code. 

For peta-scale machines, currently being deployed in the US with tens or hundreds of thousands of processors, dynamic and real-time profiling will be essential, and in particular profiling which is inherently tied into the application and automatically generated with little overhead during an application run. Current projects, in the D-Grid and US, are developing technologies for Cactus simulations to 
automatically produce and store profiling and application metadata from simulations. This information
will then be used for analysis to lead to optimized codes and potentially improved parallel computing paradigms. 



\section{Experiment and Analysis}
\label{ea}

To illustrate the advantages  of application-side adaptivity using real-time profiling 
a series of experiments was performed using the adaptive checkpointing described in Sec.~\ref{ac}. The application code used was the \codename{Ccatie} astrophysics
code~\cite{Alcubierre99d},
which can simulate the collision
of black holes, to test the adaptive checkpointing functionality.
This compute and data intensive application solves Einstein's
equations for general relativity in 3D, evolving over twenty
partial differential equations using high order finite differences.
We used the Carpet driver~\cite{Schnetter-etal-03b,carpet_web} to
provide adaptive mesh refinement.  Starting from a uniform grid with
$40^3$ grid points, we added additional refined levels every $5120$
iterations.  Such a strategy is e.g.\ necessary to simulate the
collapse of a stellar core in a supernova, where the central density
increases considerably during the collapse.  Note that for $L$
refinement levels, the computing time per iteration grows as $O(2^L)$,
while the amount of data to be checkpointed grows at the same time as
$O(L)$.

\begin{figure}[tbp]
  \includegraphics[width=0.49\textwidth]{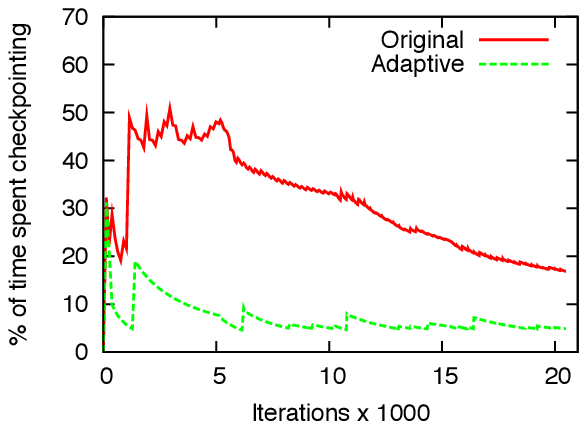}
  \hfill
  \includegraphics[width=0.49\textwidth]{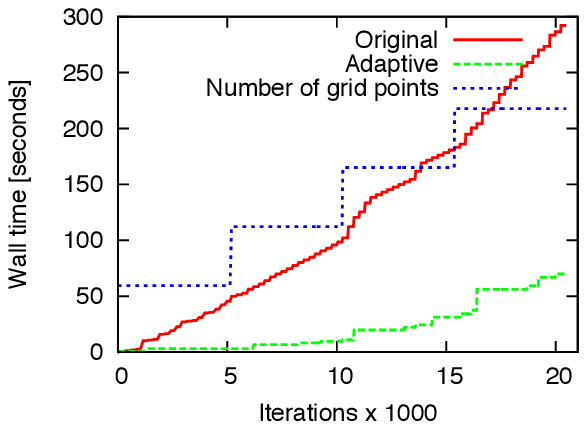}

  \caption{\emph{Left:} Percentage of time spent checkpointing during
    a run. The adaptive run keeps within the desired bound (5\%). 
    \emph{Right:}
    Total time spent checkpointing during a run.  The adaptive version
    checkpoints less frequently to keep close to the $5\%$ bound.  The
    dashed line indicates the increase in the number of grid points as
    new refinement levels are added.}


  \label{fig:results}

\end{figure}

The simulation runs for $1,737$ seconds, spending $19\%$ of the total time
checkpointing.  The original configuration checkpointed every
$512$ iterations.  With the \codename{AdaptCheck} thorn  the maximum
percentage of time spent checkpointing was restricted to $5\%$ of the total wall time
in the adaptive configuration.
Figure~\ref{fig:results} compares the actual percentage of time spent
checkpointing for the original and the adaptive configuration.  The
results show that we are able to successfully bound this measure.
This also yielded a $17\%$ reduction in the total runtime.



Each regridding increases the problem size  by $40^3$ grid points.  This
increase means that, with additional levels, each iteration takes more time
to compute, the time between periodic checkpoints will 
increase, and 
the amount of time to checkpoint increases
during the run.

A common practice is to choose the checkpointing interval to be 
 short enough at all times of the
run.  Unfortunately, this means that checkpointing occurs much too
frequently early in the run.
In another run using the \codename{AdaptCheck} thorn, we bounded the interval
between checkpoints independent of the performance of the run and the
I/O system.  This reduced the amount of time spent checkpointing from
$319\,$s to $75\,$s; the total runtime was reduced by $20\%$.
This functionality can be used to guarantee a certain level of fault
tolerance when adapting the checkpointing based on the simulation's
characteristics.

\section{Related Work and Conclusions}
\label{rwc}

The above results indicate how a scientific code can use a generic, self contained timing infrastructure for runtime profiling and adaption leading to significantly improved overall performance --- 
in this case increasing the time spent in computation and the fault tolerance of the run, while reducing  checkpointing time. The timing infrastructure  was implemented in a highly portable manner in the Cactus Framework, and is easily available to users, either via parameter choices for higher level tools, or through an API for code developers. The infrastructure is able to use 
platform dependent clocks, as well as libraries such as PAPI\@. 

A substantial amount of work has recently been seen in  automated application profiling and adaption, motivated by new possibilities in Grid computing, and  a growing realization of the new tools needed for peta-scale computing. Attention has focused on developing general libraries and tools for application profiling, adaption and steering (e.g.\ SciRun, GrADS, RealityGrid). 
For example, in the GrADS project, a program development 
framework has been developed which can encapsulate general applications as configurable object programs, and then optimize these for execution on a specific set of Grid resources~\cite{CS_Berman05a}. GrADS uses the Autopilot system for real-time application monitoring and closed loop control. Autopilot sensors can be embedded in application code by developers, or as 
in the GrADS system an automated mechanism can be used.

The Cactus timing infrastructure incorporates its own application profiling, adaption and steering. 
The 
design of the Cactus Framework also allows thorns to be easily written to connect to external packages
when these provide an advantage, as in experiments with GrADS, Autopilot, and ongoing work with SciRun. A key advantage of the Cactus infrastructure, however, is that 
there is
an intimate connection with the scientific application --- {\it even with no attention to application profiling}. Cactus applications are automatically enabled with steerable parameters and profiling at the level of thorn methods, thorns, schedule bins, as well as communication times and I/O times. 
Such an understanding of the application structure and scientific content is crucial for effective steering and control~\cite{CS_Vetter99a}.
Higher level Cactus tools 
can build on these capabilities, and leverage current work on intelligent adaption in distributed environments (e.g.\ \cite{CS_Reed05a}), to provide powerful capabilities for analysis and control of 
scientific applications in HPC and Grid environments.

\section*{Acknowledgements}

We acknowledge contributions from the Cactus Team in the
timing infrastructure implementation, in particular David
Rideout, Thomas Schweizer, John Shalf, Jonathan Thornburg, Andre
Werthmann, and Steve White.  We thank Ed Seidel for suggestions, and
Elena Caraba for her help with
preparing this manuscript.  This work was partly supported by NSF
Grant 540179 (DynaCode) and  the German Federal Ministry of Education and
  Research (D-Grid  01AK804). Computing resources were provided by the
Center for Computation \& Technology at LSU\@.



\end{document}